# Photo-enhanced metastable c-axis electrodynamics in stripe ordered cuprate $La_{1.885}Ba_{0.115}CuO_4$

K. A. Cremin[a*], J. Zhang[a*], C. C. Homes[b], G. D. Gu[b], Z. Sun[a,c], M. M. Fogler[a], A. J. Millis[c], D. N. Basov[c], R. D. Averitt[a§]

[a]Department of Physics, University of California San Diego, La Jolla, CA 92093
[b]Condensed Matter Physics and Materials Science Department, Brookhaven National Laboratory, Upton, New York 11973
[c]Department of Physics, Columbia University, New York, New York 10027

Quantum materials are amenable to non-equilibrium manipulation with light, enabling modification and control of macroscopic properties. Light-based augmentation of superconductivity is particularly intriguing. Copper-oxide superconductors exhibit complex interplay between spin order, charge order and superconductivity, offering the prospect of enhanced coherence by altering the balance between competing orders. We utilize terahertz time domain spectroscopy to monitor the c-axis Josephson Plasma Resonance (JPR) in $La_{2-x}Ba_xCuO_4$ (x = 0.115) as a direct probe of superconductivity dynamics following excitation with near infrared pulses. Starting from the superconducting state, c-axis polarized excitation with a fluence of 100 $\mu J/cm^2$ results in an increase of the far-infrared spectral weight by more than an order of magnitude as evidenced by a blueshift of the JPR, interpreted as resulting from non-thermal collapse of the charge order. The photo-induced signal persists well beyond our measurement window of 300 ps and exhibits signatures of spatial inhomogeneity. The electrodynamic response of this new metastable state is consistent with enhanced superconducting fluctuations. Our results reveal that $La_{2-x}Ba_xCuO_4$ is highly sensitive to non-equilibrium excitation over a wide fluence range, providing an unambiguous example of photo-induced modification of order-parameter competition.

\* Equal contributions

§ email: raveritt@ucsd.edu

High-temperature superconductivity in the cuprates can coexist or compete with a multitude of other phenomena, including charge and spin order, pair density waves, and the pseudogap[1–4] all of which have observable signatures in the linear or transient electrodynamic response[5]. A key question for the cuprates is the extent to which the underlying interactions can be manipulated to alter the macroscopic properties. Ultrafast pump probe spectroscopy provides a unique means to initiate and interrogate non-equilibrium dynamics and property control in superconductors[6,7].

Evidence of transiently enhanced interlayer tunneling has been reported in several cuprates following selective phonon pumping[8–10] or intense near infrared excitation[11]. In these experiments, the c-axis Josephson plasma resonance (JPR) serves as a reporter of the interlayer tunneling which, in general, scales with the superfluid spectral weight[12]. Upon applying intense near infrared pump excitation starting above the transition temperature $T_c$, Nicoletti et. al. observed a blueshift of the plasma resonance relative to the below $T_c$ equilibrium response. The plasma resonance decayed after several picoseconds and was interpreted in terms of transient superconductivity[11]. Similar dynamics were subsequently observed in both $La_{2-x}Ba_xCuO_4$ and $YBa_2Cu_3O_{6+x}$ over a range of doping levels and excitation conditions. More recently, in $La_{2-x}Ba_xCuO_4$ (x = 0.095), a longer lived (> 50 ps) collective response was observed subsequent to photoexcitation at $T<T_c$ in which the JPR appeared to split into two distinct longitudinal modes[13], reminiscent of the static electrodynamic response in bilayer cuprates[14–16]. Further no pump-induced effect was observed above $T_c$. Notably, the x = 0.095 material does not exhibit robust charge or stripe order[17].

In this report, we clarify the details of the photoinduced order parameter control in the cuprates, by performing temperature and fluence dependent c-axis measurements on materials with equilibrium signatures of phase competition. We present near-infrared pump, THz probe experiments of $La_{2-x}Ba_xCuO_4$ (x = 0.115), for which we observe distinct dynamics above and below $T_c$. The doping x = 0.115 is close to the anomalous x = 1/8 composition where 3D superconductivity is maximally suppressed by robust charge and stripe order[2,18–20]. At x = 0.115, charge order, spin order and 3D superconductivity onset at $T_{co}$ = 53 K, $T_{so}$ = 40 K, and $T_c$ = 13, respectively[17]. In this compound $T_c$ is high enough to enable initiating dynamics from *within* the superconducting state. The $La_{1.885}Ba_{0.115}CuO_4$ crystal was cut and polished to expose the *a-c* plane with a large area of 5 mm x 5 mm. Temperature dependent FTIR measurements were carefully performed to provide a baseline static characterization of the electrodynamic response (Supplementary S1).

The static and dynamic reflectivity of LBCO was measured as a function of temperature and fluence using terahertz time-domain spectroscopy (0.15-2 THz) (Supplementary S1). As shown in Fig. 1(a), both the near-infrared pump (1.55 eV) and the THz probe pulses are polarized along the c-axis. The static THz reflectivity is plotted in Fig. 1(b), showing a flat response at 30 K (grey line, with a slight upturn below 0.5 THz). In the superconducting state (7K, blue curve) a sharp reflectivity edge emerges around 200 GHz, a hallmark of the inter-layer JPR effect in agreement with previous studies for this doping[11].

For the dynamics measurements, we first performed one-dimensional scans, where the photoinduced change in the peak electric field of the single-cycle THz pulse ($\Delta E/E$) is measured as a function of pump-probe delay. This raw unprocessed data unambiguously highlights important features of the electrodynamic response. Figure 1(c) plots one-dimensional scans ($\Delta E/E$ peak scans) for various fluences, starting from an initial temperature of 7K, well into the superconducting state. The dynamics exhibit a long-lived response with an initial risetime of several picoseconds. The magnitude of the $\Delta E/E$ signal slightly increases from 50 to 100 $\mu J/cm^2$, followed by a strong *decrease* in amplitude at higher fluences up to 760 $\mu J/cm^2$. The observed fluence dependence indicates dynamics that are distinct from a photoinduced decrease in the condensate density with a commensurate increase in the quasiparticle density. In that scenario, $\Delta E/E$ would increase in magnitude with increasing fluence.

Calculations using the two-temperature model of the initial electron-phonon (e-ph) thermalization (Supplementary S3) indicate a rise in the final temperature that increases with fluence (after several picoseconds when the plateau in the data in Fig. 1(c) is reached). Figure 1(d) displays the phase diagram of LBCO (reproduced from Ref. [17]) with color coded dots (corresponding to the fluences used in Fig. 1(c)) indicating the final temperature after thermalization. Clearly, the amplitude of $\Delta E/E$ decreases with increasing fluence as $T_{co}$ is approached after the initial e-ph thermalization. This indicates that the strongest photo-induced response occurs for fluences that do not heat the sample above $T_{co}$. We refer to 50 – 380 µJ/cm² as the low fluence regime where the quasi-equilibrium temperature stays below $T_{co}$. The data in Figs. 2 and 3 reveal that the dynamics are dramatically different for low- and high-fluence (>380 µJ/cm²) regimes spanning $T_{co}$. The difference in dynamics is evident in Fig. 1(c), where the 760 µJ/cm² data exhibits an exponential decay in contrast to the plateau apparent in the low fluence data.

We now discuss the full spectroscopic response of the dynamics, considering first low fluence optical excitation of LBCO (100 µJ/cm²) obtained starting from an initial temperature of 7 K. The optical properties of the photo-excited region were extracted using a layered model to account for the penetration depth mismatch between the pump and probe beams (Supplementary S2). Figure 2(a) shows the photoinduced reflectivity (blue curve at 10 ps, red curve at 300 ps), revealing a large reflectivity increase extending out to 1 THz, corresponding to an increase in the plasma frequency from ~0.2 to 0.9 THz. This is more clearly revealed in Figs. 2(b) and (c) showing the loss function -Im(1/$\varepsilon$) (where $\varepsilon = \varepsilon_1 + i\varepsilon_2$ is the c-axis dielectric response) which peaks at the plasma frequency. Importantly, there is no sign of decay of the photo-induced state over the measured temporal window, indicating a metastable state that persists beyond 300 ps. Figure 2(d) and (e) show the associated c-axis optical conductivity ($\sigma = \sigma_1 + i\sigma_2$) highlighting an important observation. Namely, there is a peak in $\sigma_1$ at 0.3 THz signifying dissipation in the c-axis THz transport. In contrast, in equilibrium for a superconductor, $\sigma_1$ at frequencies greater than zero but less than twice the superconducting gap approaches zero (solid grey line in Fig. 2(d)). This origin of the peak in $\sigma_1$ arising after photoexcitation will be discussed below. Finally, the solid green line in Fig. 2(a) shows the spectral response at 10 ps delay for a fluence of 100 $\mu J/cm^2$ taken at an initial temperature of 20K – that is, above $T_c$. There is an increase in the reflectivity, arguably with the development of a weak plasma edge. However, as shown in Fig. 2(f), the response is much smaller and shorter lived in comparison to dynamics initiated from within the superconducting state. In short, the emergence of a robust metastable state upon low fluence photo-excitation requires starting from the superconducting state while remaining below $T_{co}$ after e-ph thermalization.

To complete the data discussion, we now consider the high fluence dynamics. Figure 3 shows the spectroscopic results for high fluence excitation at 9 $mJ/cm^2$ which leads to a final temperature greater than $T_{co}$. Fig. 3(a) reveals an increased reflectivity starting from an initial temperature of 30 K. At early times a broad peak emerges in the loss function (shown in Fig. 3(b) and (c)) that rapidly broadens and decays on a ~10 ps timescale. There is a corresponding increase in $\sigma_1$ and $\sigma_2$ (Fig. 3(d) and (e)), though there are no well-defined peaks as for the low fluence results. Fig. 3(f) reveals the rapid decay in the transient $\Delta E/E$ with 9 $mJ/cm^2$ excitation, with the inset showing a saturation of the dynamics with increasing fluence. The plateau in the $\Delta E/E$ scans following the initial exponential decay is presumably associated with heating, leading to a broad and featureless optical conductivity from 0.1 – 1.5 THz. The broadened plasma edge that appears after the arrival of the pump pulse is qualitatively in agreement with the previous study by Nicoletti et. al.[11]. Given the relatively small value of $\sigma_1$, this could indicate the presence of enhanced superconducting correlations associated with the unambiguous transient blueshift of the plasma frequency. Summarizing, the high fluence transient response evolves at temperatures above $T_{co}$ and is short lived, in contrast to the low fluence metastable results presented in Fig. 2, implying that the two features correspond to distinct states.

We now discuss the origin and nature of the low fluence photoinduced metastable electrodynamic response (Figure 2). For 100 µJ/cm$^2$, the temperature following e-ph thermalization is ~35 K. In equilibrium, 35 K is within the charge ordered, non-superconducting region that is spectrally featureless at THz frequencies. In contrast, the low fluence metastable state produced by c-axis excitation exhibits unique spectroscopic features in the loss function -Im(1/$\varepsilon$) (Fig. 2(b) and (c)) and conductivity (Fig. 2(d) and (e)) that are indicative of a non-equilibrium metastable state. Importantly, the plasma frequency ($\omega_p$) is blueshifted by a factor of 4.5 upon photoexcitation (from 0.2 → 0.9 THz) corresponding to an increase of the low energy spectral weight ($\Sigma$) by a factor of 20 (i.e. $\Sigma \propto \omega_p^2$). This is a robust observation, independent of the microscopic origin of the electrodynamic response. Additionally, the spectral weight must originate from energies beyond the 1.5 THz regime probed here because the equilibrium condensate spectral weight (Fig. 4(a)) is far too small to account for the factor of 20 increase. A change in the c-axis response may also arise from structural distortions affecting interlayer coupling independent of a change in the in-plane superfluid density[21]. Although our measurements are not directly sensitive to structural distortions, we estimate the metastable phase to be at ~35 K which is still below the low temperature orthorhombic to tetragonal transition occurring at 53 K[22]. As such, this is not likely to be the cause of the increase in $\Sigma$. We next show that a more likely scenario for the increase in $\Sigma$ is photoinduced collapse of charge/stripe order whereby spectral weight originally at energies above the CDW gap scale is made available for enhanced c-axis transport.

A well-known approximate scaling relationship (valid for both c-axis and ab-plane) for the cuprates is $\rho_o = 120\sigma_{dc}T_c$, where $\rho_o$ is the superfluid density ($\propto \omega_p^2$) and $\sigma_{dc}$ is the zero frequency conductivity measured just above $T_c$[23]. Figure 4(a) shows the product $\sigma_{dc}T_c$ versus $\rho_o$ for x = 0.115 doping measured in this study, and for x = 0.95 and x = 0.145 taken taken from Ref. [22] For x = 0.115 at 7 K, the JPR appears near the renormalized superconducting plasma frequency $\omega'_p$ = 6.7 cm$^{-1}$ (200 GHz). The unscreened plasma frequency is given by $\omega_p = \omega'_p\sqrt{\varepsilon_\infty}$ ($\varepsilon_\infty \sim 30$) giving a superfluid density $\rho_o \equiv \omega_p^2$ = 1.3 x 10$^3$ cm$^{-2}$ (This is taken to be very near the superfluid density at T = 0 K). The conductivity just above $T_c$ = 13 K is taken as $\sigma_{dc} \sim 0.8$ $\Omega^{-1}$cm$^{-1}$, estimated from transport measurements on a similar doping of x = 0.11[24]. As seen in Fig. 4(a), the x = 0.115 doping is consistent with $\rho_o = 120\sigma_{dc}T_c$ plotted as a dashed line. We now extend this concept to the photoinduced metastable state which, as described above, exhibits a large increase in spectral weight.

From the photoinduced blueshift of the plasma edge ($\omega'_e \sim 0.9$ THz or 30 cm$^{-1}$) it is possible to determine the photoinduced density $\rho_e$. This estimate from experimental data gives $\rho_e$ = 2.7 x 10$^4$ cm$^{-2}$, yielding $\rho_e/\rho_o \sim 20$ as mentioned above. We now assume the validity of the scaling relation $\rho_e = 120\sigma_{co}T_{co}$ where $\sigma_{co} \sim 3$ $\Omega^{-1}$cm$^{-1}$ is the conductivity just above the charge ordering temperature ($T_{co}$ = 53 K). From this we obtain $\rho_e/\rho_o \sim \sigma_{co}T_{co}/\sigma_c T_c \sim$ 15, in reasonable agreement with experiment. The product $\sigma_{co}T_{co}$ is plotted in Fig. 4(a) as a red data point and lies on the dashed line given by $\rho_s$ = 120$\sigma_{dc}T_c$. This suggests that c-axis spectral weight initially tied up in charge order is released upon photoexcitation leading to the blueshift of the plasma edge. We note that the precise microscopic reason for the photoinduced collapse of charge order upon c-axis interband excitation is not understood and is a topic for future investigation, beyond the scope of the present study.

The peak in the loss function cannot be simply described as an enhanced plasma edge with increased interlayer tunneling arising from the collapse of the charge order. This is because we have a peak in $\sigma_1$ at a non-zero frequency (Fig. 2(d)). In principle, two inequivalent junctions (with different inter-layer spacing) do yield two distinct longitudinal JPRs at different frequencies. For example, Sm doped La$_{2-x}$Sr$_x$CuO$_4$ (LSCO), is composed of a stack of inequivalent Josephson junctions. Out of phase oscillations of the two longitudinal modes result in a transverse mode with a resonance frequency intermediate to the two JPRs with a corresponding peak in $\sigma_1$[16,25]. In such a case, the spectral weight would be associated with a pure superfluid response, which is certainly intriguing since our LCBO sample is (as discussed above) at ~35 K following e-ph equilibration (i.e. ~2.7$T_c$). It is, however, not clear how photoexcitation could lead to uniform creation of a well-defined microscopic bi-layer structure, though spectroscopic aspects of this are present in the non-charge ordered LBCO (x = 0.095)[13]. However, for

x = 0.115, charge order collapse plays a dominant role as described above. As such, we consider an alternative scenario based on photoinduced mesoscopic inhomogeneity.

The simplest effective medium theory for an anisotropic layered crystal (appropriate to c-axis cuprates) is schematically depicted in Fig. 4(b). The effective dielectric response $\varepsilon_{eff}$ is given as $1/\varepsilon_{eff} = f_1/\varepsilon_1 + f_2/\varepsilon_2$ where $f_1$ and $f_2$ are the volume fractions corresponding to regions with complex dielectric function $\varepsilon_1$ and $\varepsilon_2$, respectively[26]. Taking $f_1$ as the superconducting volume fraction (having the 7K equilibrium response), and $f_2$ as a Drude response (with a finite scattering rate of 0.36 THz) yields the fit to the experimental data shown in Fig. 4(c) (additional details presented in the SI). Notably, the peak in $\sigma_1$ corresponding to a photoinduced transverse mode is accurately reproduced with this model taking $f_1 = 0.02$ and $f_2 = 0.98$. While the general features of $\sigma_2$ are reproduced by the effective medium model, an exact fit was not possible. We note that errors in extracting $\sigma_2$ from experiment are generally more difficult in comparison to $\sigma_1$ as this depends sensitively on changes in phase of the terahertz pulse.

Taking the fits as representative of the photoinduced state leads to some interesting conclusions. First, the preponderant component consisting of a Drude response is consistent with the picture presented above that the pump destroys the charge/spin order. Second, a small but non-negligible superconducting volume fraction (~2%) is required to obtain the spectral response in Fig. 4(c). This indicates that even though the temperature is more than two times greater than the equilibrium $T_c$, regions of superconductivity persist. Furthermore, the Drude response (associated with $f_2$) is anomalous as it exhibits a small scattering rate of 0.36 THz. This is surprising for a non-superconducting c-axis response and has not been observed in the equilibrium c-axis response for any cuprate material. We suggest that the origin of this enhanced response arises from incipient superconducting correlations that enhance the c-axis conductivity as has been theoretically discussed [27,28]. This is consistent with the experimental observation that the observed metastability requires starting from an initial superconducting state and that a non-zero superconducting volume fraction persists after photoexcitation.

In summary, low fluence photo-excitation favors the establishment of a novel long lived state with superconductivity playing an important role. Additional experiments are required to fully characterize the observed metastable response and the effective medium description of mesoscale inhomogeneity that includes a superfluid response. Our results raise crucial theoretical questions including the origin of the superconductivity and the physics and surprisingly long lifetime of the metallic state. The observed long lifetime bodes well for performing additional experiments, including time-resolved nanoscopy to spatially resolve the photoinduced state.

**Acknowledgements:** Funding DOE-BES **DE-SC0018218**. Work at Brookhaven National Laboratory was supported by the Office of Science, U.S. Department of Energy under Contract No. **DE-SC0012704**. We thank Danielle Nicoletti & Andrea Cavalleri as well as Ryo Shimano & Rysuke Matsunaga for fruitful discussions.

**Data availability:** The data that support the findings of this study are available from the authors on reasonable request.

Figure 1

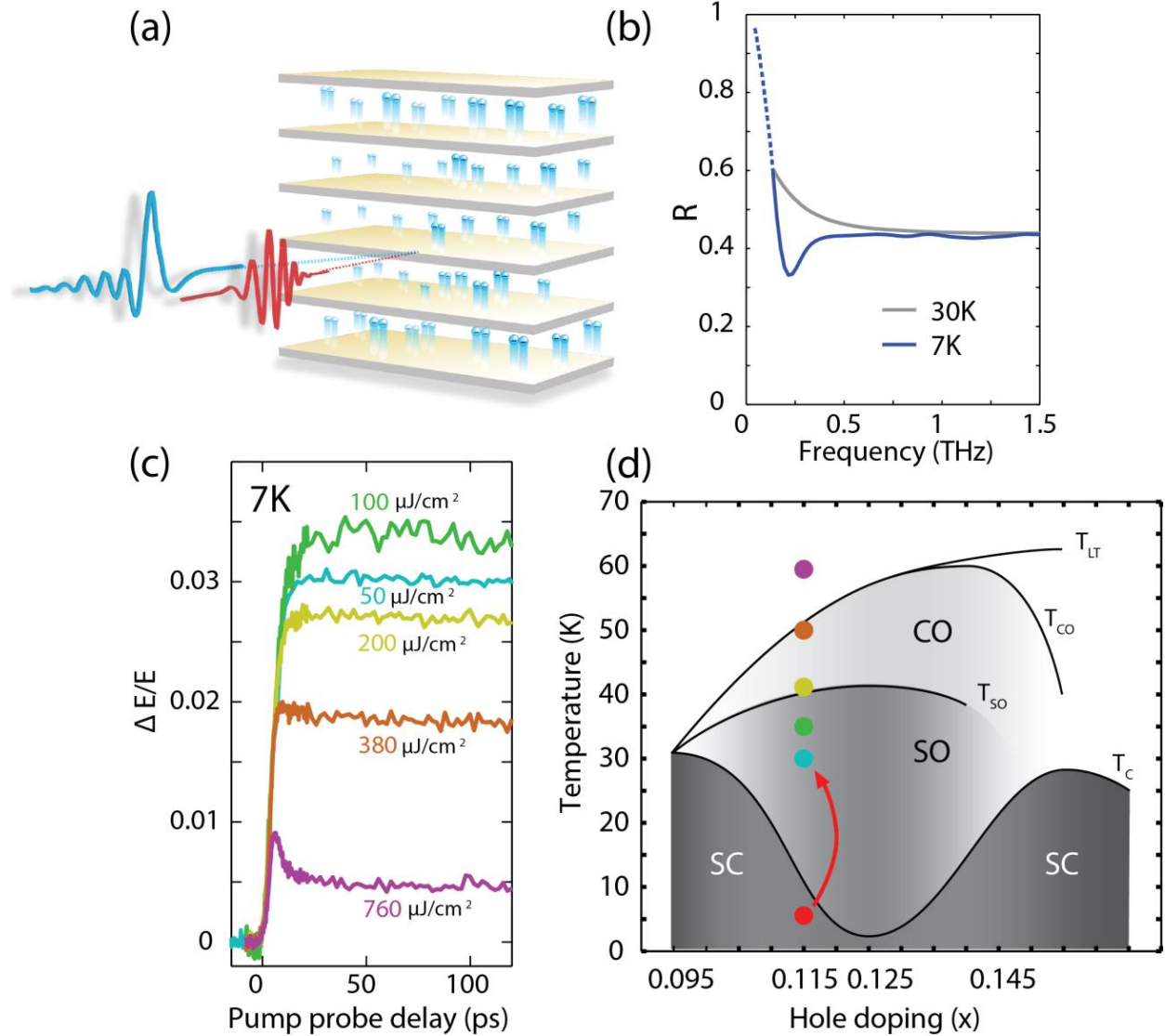

Figure 1: (a) Schematic of the 1.55 eV pump THz probe experiment on LBCO crystal with pump and probe polarization along c-axis. The illustration depicts the superconducting state with electron pairs tunneling along the c-axis. (b) Equilibrium THz reflectivity above and below $T_C$ at 30 K and 7 K, respectively. The dotted line is beyond our experimentally accessible spectral range and is a guide to the eye. (c) Time dependent relative change in the THz electric field amplitude after excitation at various pump fluences (at 7 K). (d) Temperature versus hole doping phase diagram of $L_{2-x}Ba_xCuO_4$ (reproduced from Ref. [17]). Regions of the phase diagram include bulk superconductivity (SC) at onset temperature $T_c$, spin ordering (SO) at temperature $T_{so}$, charge ordering at $T_{co}$, and low-temperature structural transition $T_{LT}$. The initial temperature at 7 K for x = 0.115 is plotted in red, and the color dots mark the estimated lattice temperature after pump excitation and e-ph thermalization has occurred (colors corresponding to pump fluences shown in panel (c)).

Figure 2

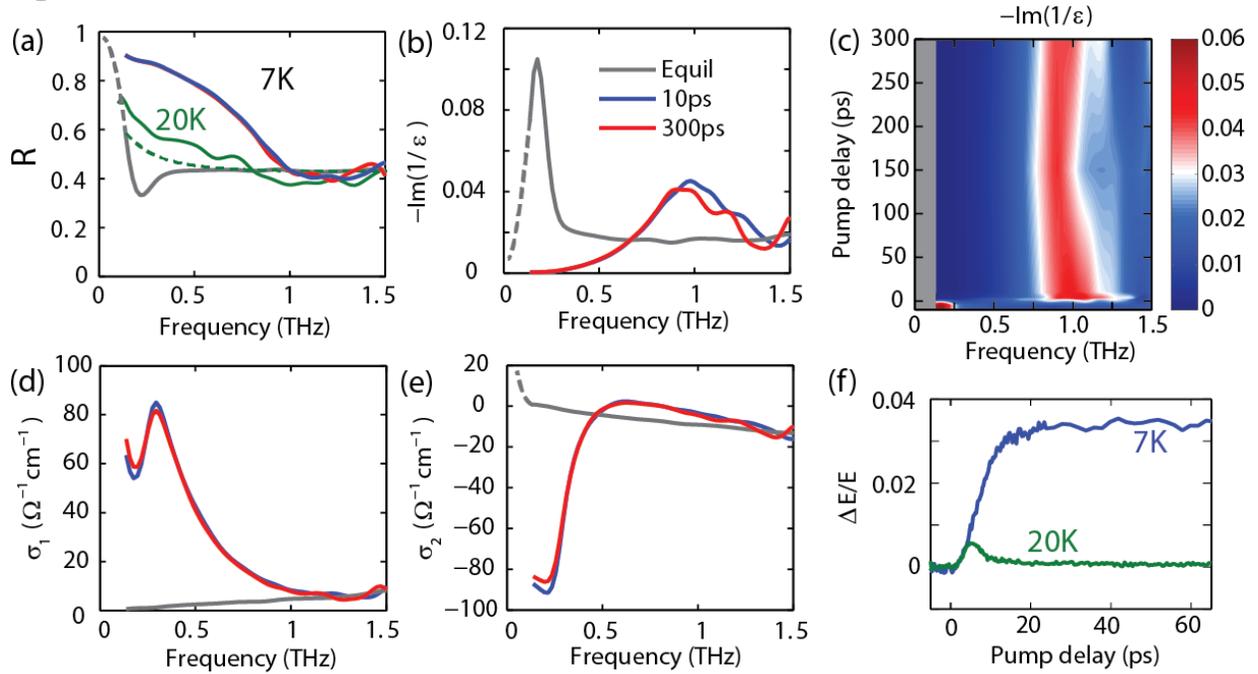

Figure 2: Extracted c-axis THz optical properties of LBCO at different pump-probe delays after photo-excitation (colored) with 100 µJ/cm$^2$ and at equilibrium (grey). All data has been taken at 7 K below $T_C$ except for the green curves in panels (a) and (f) which were taken at 20 K. (a) Reflectivity at 7 K before (grey) and after photo-excitation (colored) at different pump-probe delays. Plotted in green is the equilibrium (dotted) reflectivity and largest photo-induced change (solid) in reflectivity at 20 K. (b) Loss function -Im(1/ε). Dashed grey line is beyond our spectral resolution and is a guide to the eye. (c) Spectral evolution of the loss function after photo-excitation. (d)-(e) Real and imaginary parts of the THz conductivity. (f) Peak of ΔE/E THz transient after photo-excitation at 7 K and 20 K.

Figure 3

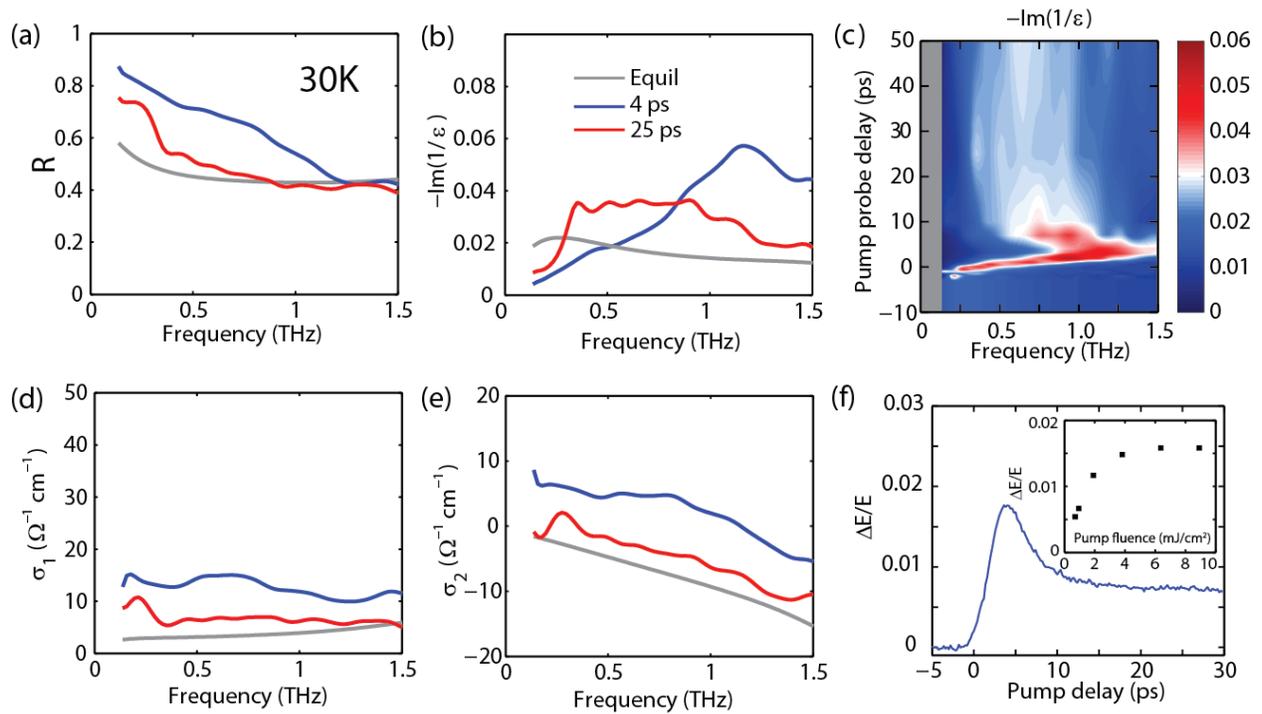

Figure 3: Extracted c-axis THz optical properties of LBCO at different pump-probe delays after photo-excitation (colored) with 9 mJ/cm$^2$ and at equilibrium (grey). All data taken at 30 K above $T_C$ and below $T_{CO}$. (a) Reflectivity and (b) loss function, -Im(1/ε) at different pump probe delays. (c) Spectral evolution of the loss function after photo-excitation. (d)-(e) Real and imaginary parts of the THz conductivity at different pump probe delays. (f) Time dependent relative changes in THz electric field after photo-excitation. The inset displays the maximum ΔE/E value at 30 K as a function of pump fluence.

Figure 4

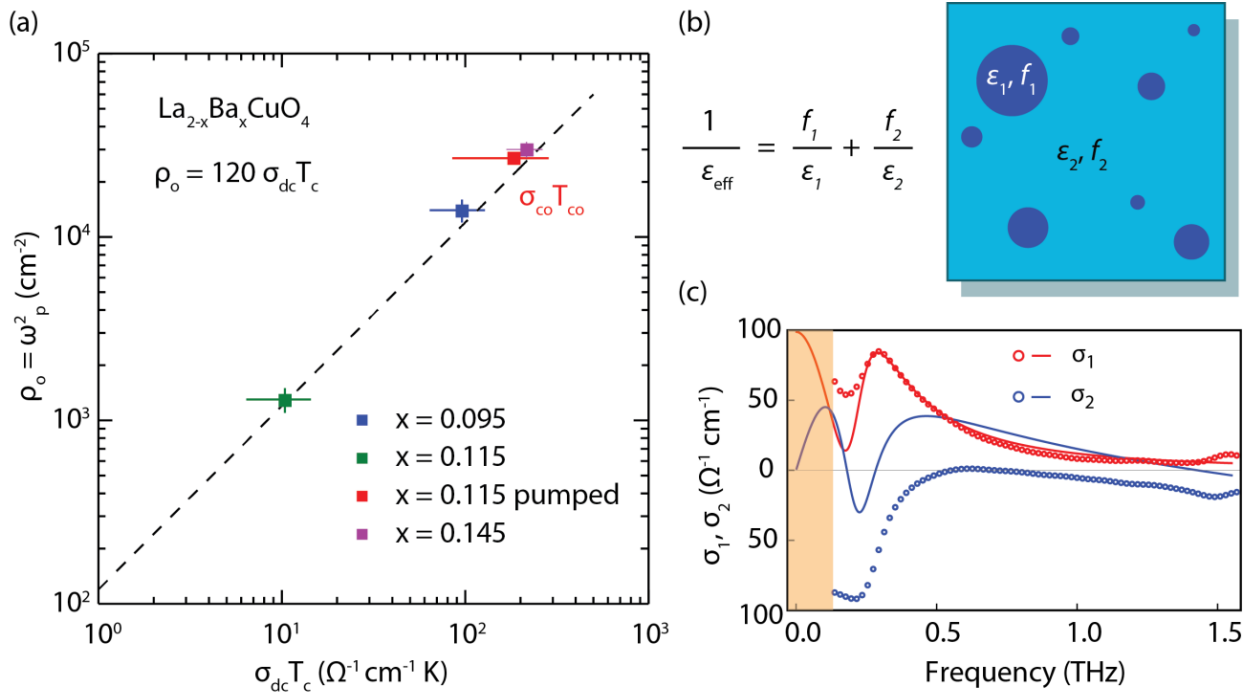

Figure 4: (a) Plot of the superfluild density $\rho_s$ versus the product of the zero frequency conductivity ($\sigma_{dc}$) and the superconducting transition temperature $T_c$ for several dopings of LBCO ($\sigma_{dc}$ is the conductivity measured just above the superconducting transition). The values for x = 0.95 and x = 0.145 were taken from Ref. [22]. The dashed line is the universal scaling relation for cuprates $\rho_s = 120\sigma_{dc}T_c$ found by Homes et. al.[23] (b) Schematic of anisotropic effective medium model with different dielectric constants $\varepsilon_1$ and $\varepsilon_2$ and filling fractions $f_1$ and $f_2$ respectively. The superconducting volume is depicted with dark blue circles and the excited region is in light blue. (c) Real (red) and imaginary (blue) parts of the THz conductivity after photo-excitation with 100 µJ/cm$^2$ at 7 K. Experimental data is plotted with dots and the effective medium model with solid lines.

# Supplementary Materials
## Photo-enhanced metastable c-axis electrodynamics in stripe ordered cuprate La$_{1.885}$Ba$_{0.115}$CuO$_4$

K. A. Cremin[1], J. Zhang[1], C. C. Homes[2], G. D. Gu[2], Z. Sun[1], M. M. Fogler[1], D. N. Basov[3], Richard Averitt[1]

[1]Department of Physics, University of California San Diego, La Jolla, CA 92093
[2]Condensed Matter Physics and Materials Science Department, Brookhaven National Laboratory, Upton, New York 11973
[3]Department of Physics, Columbia University, New York, New York 10027

**Figures:**

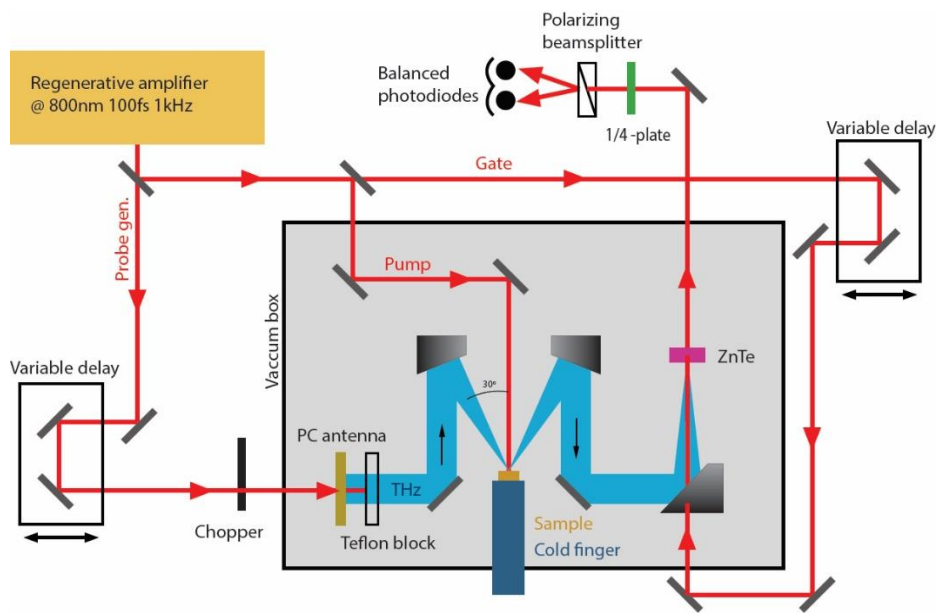

Figure S1: Schematic of experimental setup for near-IR pump THz probe.

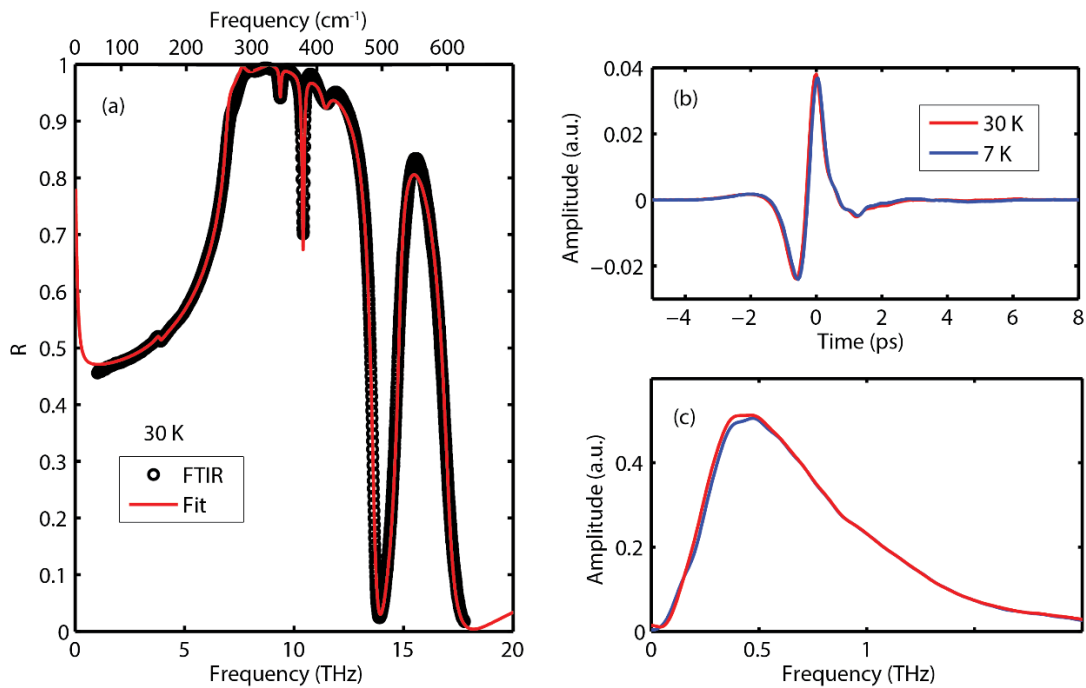

Figure S2: (a) c-axis reflectivity FTIR data taken at 30K with fit. (b) Time domain profile of THz pulse reflected from the sample at 30 K and 7 K. (c) Fourier transform of the time domain pulses in panel (b).

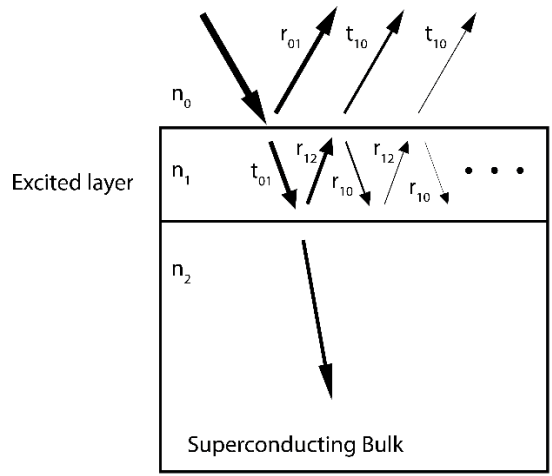

Figure S3: Schematic drawing of thin layer model. In this model the thin excited layer sits on top of the bulk crystal containing the equilibrium optical properties.

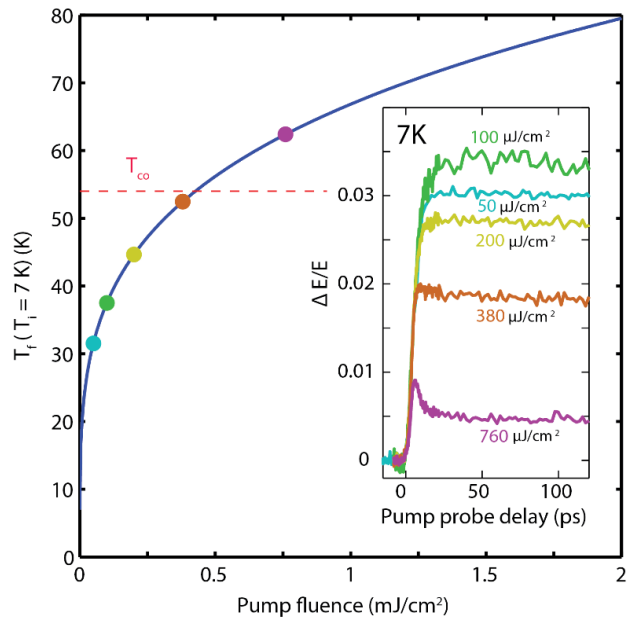

Figure S4: Calculated final lattice temperature after electron-phonon thermalization. The colored dots represent $T_f$ for a given fluence which are color coordinated with the inset figure. The inset is the time dependent relative change in the THz electric field amplitude after excitation at various pump fluences (at 7 K). The red dashed line is the charge order transition temperature $T_{co}$.

## S1: Methods

The La$_{1.885}$Ba$_{0.115}$CuO$_4$ (LBCO) crystal studied was cut and polished to expose the *ac* surface with a large area of 5 mm x 5 mm. The c-axis properties were probed with broadband THz pulses generated from a commercial GaAs based photo-conductive antenna using incident light from a 1 kHz Ti:Sapphire regenerative amplifier. The THz pulses produced have a usable bandwidth from 0.15-2 THz which allowed a measurement of the equilibrium Josephson plasma resonance (JPR) near 0.2 THz for x = 0.115 doping. Figure S1 shows a schematic of the optical pump-THz probe measurement in reflection.

The probe pulses were focused onto the sample at an incident angle of 30 degrees with the electric field polarized along the c-axis. The THz pulses were collected after the sample and measured via electro-optic sampling with an 800 nm gate pulse in a 2mm thick ZnTe crystal. The LBCO crystal was photoexcited with 100 fs, 800 nm wavelength pulses, polarized along the c-axis with a beam diameter of 6 mm FWHM to ensure a uniform excitation across the sampled region.

The equilibrium c-axis reflectivity above T$_c$ was characterized with broadband FTIR measurements at 30 K which is displayed in Fig. S2. The equilibrium complex index of refraction n($\omega$) was determined by fitting the reflectivity curve with Drude-Lorentz oscillators placed at known IR phonon frequencies[1]. The fit was accomplished using the software package RefFIt$^{TM}$ and is shown in Fig. S2. A small Drude component was added to give a dc conductivity $\sigma_1(\omega=0)$ ~ 3 $\Omega^{-1}$cm$^{-1}$ at 30 K to match previous transport measurements[1,2].

The reflection coefficient of a material is given by $r(\omega) = E_s(\omega)/E_i(\omega)$ where $E_s(\omega)$ is the electric field reflected off the sample and $E_i(\omega)$ is the incident field. By measuring the THz waveform $E_s(t)$ via electro-optic sampling, we take the Fourier transform to obtain $E_s(\omega)$. The reflection coefficient was then determined at 7 K by measuring $E_{s,7K}(\omega)$ and using the relation

$$r_{7K} = \frac{E_{s,7K}(\omega)}{E_{s,30K}(\omega)} r_{30K} \qquad (1)$$

where $r_{30K}$ is the reflection coefficient obtained from the Drude-Lorentz fitting. Figure S2(b)(c) displays a measurement of the single cycle THz pulses reflected from the sample at 30 K and 7 K in the time domain and frequency domain respectively.

## S2: Layer Model and parameter extraction

The photo-induced change $\Delta E_s(t,\tau)$ in the reflected electric field was measured at various pump probe time delays $\tau$, over the temporal window $t$ of the THz pulse. The quantity $\Delta E_s(t,\tau) = E_{s,pumped}(t,\tau) - E_{s,unpumed}(t,\tau)$ was acquired at each delay time $\tau$ by using a lock-in triggered by the modulation of the pump pulse with a mechanical chopper. The unpumped electric field was measured 40 ps before the arrival of the pump pulse.

The raw photo-induced changes $\Delta E$ measured in the reflected field require further processing to extract the complex optical properties of the excited region. The penetration depth mismatch between 800 nm pump (~400 nm) and THz probe (~100-300 μm) results in a relatively small change in $\Delta E$ (~1-3 %) after photo-excitation and is taken into account when modeling the THz response. We model the photo-excited region as a thin layer with a thickness of $d = 400$ nm on top of an unperturbed bulk containing the material properties of the sample in equilibrium before the arrival of the pump pulse. The total complex reflection coefficient $r'(\omega)$ from the layered system is the summation of all internal reflections displayed in Fig. S3 and can be expressed in closed form as:[3]

$$r'(\omega) = \frac{r_{01} + r_{12} e^{i2\delta}}{1 + r_{01} r_{12} e^{i2\delta}}, \tag{2}$$

$$\delta = \frac{2\pi d}{\lambda} \sqrt{n_1^2 - \sin^2 \theta_1} \tag{3}$$

where $r_{01}$ and $r_{12}$ are the reflection coefficients from the front and backside of the excited layer, respectively, and $\lambda$ is the THz wavelength. The phase accumulation as the THz pulse travels through the excited layer is given by $\delta$. The complex reflection coefficient $r'(\omega,\tau)$ was determined from the experimentally measured quantities $\Delta E_s(\omega,\tau)$ and $E_s(\omega)$ using the relation $r'(\omega,\tau) = r(\omega)[\Delta E_s / E_s + 1]$ where $r(\omega)$ is the equilibrium reflection coefficient. A numerical solution to the above Fresnel equation was found, returning a value for the complex index of refraction $n_1(\omega,\tau)$ of the excited layer. We calculate the complex conductivity of the photo-excited layer

$$\sigma(\omega,\tau) = \frac{\omega}{4\pi i} [n_1(\omega,\tau)^2 - \varepsilon_\infty], \tag{4}$$

using $\varepsilon_\infty = 4.5$ as a standard value for the cuprates[4]. The photo-excited reflectivity plotted in Fig. (2) and (3) in the main text were recalculated using the complex index of refraction $n_1(\omega,\tau)$.

### S3: Pump induced heating

Heating effects are present in all pump probe measurements and the effective lattice temperature must be considered after photo-excitation. We use a two-temperature model to estimate the final lattice temperature after electron-phonon

thermalization has occurred (in ~1ps). The specific heat for $La_{2-x}Ba_xCuO_4$ is described by the relation

$$C_s = \gamma T + \beta T^3 \tag{5}$$

over a wide temperature range[5], where $\gamma$ and $\beta$ are the electronic and lattice coefficients to the specific heat, respectively. Specific heat coefficients were taken as $\gamma$ ~ 2.5 mJ mol$^{-1}$ K$^{-2}$ and $\beta$ ~ 0.25 mJ mol$^{-1}$ K$^{-4}$ estimated from literature values of nearby dopings[5]. For a given pump fluence we estimate the absorbed energy over the photo-excited region and calculate the effective temperature after electron phonon-thermalization by integrating

$$Q_{\text{pump}} = \int_{T_i}^{T_f} NC_s(T)dT, \tag{6}$$

where $Q_{\text{pump}}$ is the total energy absorbed from the pump pulse, $N$ is the number of moles in the excited volume, $T_i$ is the initial temperature and $T_f$ is the final temperature after electron-phonon thermalization. The absorbed energy $Q_{\text{pump}}$ is estimated by $Q_{\text{pump}} = F*A(1-R)$ where $F$ is the pump fluence, $A$ is the area of the pump beam on the sample using the FWHM (~6 mm) as the diameter, and $R$ is the reflectivity at 1.55 eV which is ~ 0.15[1]. The excited volume is estimated as a cylindrical disk with a diameter of 6 mm and height equal to the penetration depth.

The final temperature $T_f$ is calculated numerically and plotted versus pump fluence in Fig. S4 for an initial temperature $T_i$ = 7 K. The colored points correspond to the various fluences used in the experiment and the $\Delta E/E$ pump probe traces displayed in the inset.

**References:**
1. Homes, C. C. *et al.* Determination of the optical properties of $La_{2-x}Ba_xCuO_4$ for several dopings, including the anomalous x=1/8 phase. *Phys. Rev. B* **85**, 134510 (2012).